\documentclass[11pt, epsf]{article}
\usepackage{epsfig}
\def\ga{\mathrel{\raise.3ex\hbox{$>$\kern-.75em\lower1ex\hbox{$\sim$}}}}
\def\la{\mathrel{\raise.3ex\hbox{$<$\kern-.75em\lower1ex\hbox{$\sim$}}}}

\def\I_M{{I_{\scriptscriptstyle M\times M}}}

\begin{document}

\thispagestyle{empty}
\rightline{IP/BBSR/2004-34}

\vskip 2cm \centerline{ \Large \bf  $R$-charged AdS bubble}

\vskip .2cm

\vskip 1.2cm

\centerline{ \bf Anindya Biswas, Tanay Kr. Dey and 
Sudipta Mukherji\footnote{Electronic address: 
anindyab,tanay, mukherji@iopb.res.in}}
\vskip 10mm \centerline{ \it Institute of Physics, 
Bhubaneswar-751 005, India} 
\vskip 1.2cm
\vskip 1.2cm
\centerline{\bf Abstract}
\noindent
We construct R-charged adS bubbles in $D =5$, ${\cal N} =8$
supergravity. These bubbles are charecterised by four parameters. The asymptotic 
boundary of these solutions are deSitter times a circle. By comparing 
boundary energies, we study the possibility of a transition from certain 
class of black holes to these bubbles below a critical radius of the 
boundary circle. We argue that this may occur when  four parameters of 
the bubble satisfy a constraint among themselves. 
\newpage
\setcounter{footnote}{0}
\noindent
In this Letter, we first construct the bubble solutions of $D=5$, ${\cal 
N}=8$ gauged IIB supergravity \cite{bcs, cg} . These bubbles carry 
the R-charges that
correspond to the three $U(1)$ Cartan subalgebra of the gauge group
$SO(6)$. At asymptotic boundary, these solutions approach $S^1 \times 
dS^3$. This is similar to the adS-Schwarzschild bubbles \cite{bri, 
br}\footnote{Bubble solutions in flat space-time were constructed and 
analysed in \cite{witten, afhs}.}. We then  compute 
various components of the boundary stress-energy tensors associated with
these R-charged bubbles.
One of the main motivation for us to study these bubbles is an issue 
pointed out in \cite{st}\footnote{This was pointed to us about a year 
back by Balram Rai}. It was found earlier that a certain class of 
five dimensional black holes \cite{banadoso, banadost} (which are higher 
dimensional analogue of BTZ 
black holes) had same asymptotic boundary metric as that of 
the adS-Schwarzschild bubbles. As far as the boundary 
geometry is concerned, the only difference is that, for the black hole, 
the boundary radius of the $S^1$ (say parametrised by $\chi$) has no 
restriction. However, for the bubble, the size of $S^1$ has an upper bound (say 
$\Delta \chi_c$). Otherwise, for $\Delta\chi > \Delta\chi_c$,  the 
space-time becomes singular\footnote{
We will always assume antiperiodic boundary conditions of the fermions 
along $\chi$ for both the orbifold and the bubble.}. 
Furthermore, it was found in 
\cite{cai} 
that the components of the boundary stress tensor
of these holes are exactly same as that of a bubble in the limit of zero 
bubble radius. In general, however, the energy of the bubble space-time is 
lower than that of a black hole. Now suppose that we start with a black 
hole with large $\Delta \chi$ which parametrises the period of $S^1$ 
at the boundary. As we tune $\Delta \chi$ below the critical value $\Delta 
\chi_c$, the bubble solution for the bulk becomes available whose energy 
is less than that of a black hole. One would then expect a 
transition from the black hole to the bubble configuration. In the case of 
$R$-charged bubble, we also find a possibility of such a transition. Here 
the transition is infact from the same adS orbifold (with some radiation 
of massless matter) to R-charged bubble below certain critical radius of 
the boundary circle. In the following we construct the bubbles and then 
discuss the possibility of such a transition. In particular, this 
happens when the parameters specifying the solution satisfy certain 
relation
among themselves. Due to the conjectured dS/CFT  correspondence, one would 
then expect to see the signature of such a 
transition in the gauge theory on the boundary. One would 
also like to understand this phenomenon better in terms of string theory 
in the bulk. Neither of these issues are clear to us at this moment. We 
only offer some possibilities at the end of this Letter. 

\noindent
The bubble solution in $D=5$, ${\cal N}=8$ gauged type IIB
supergravity has the following form\footnote{This bubble metric
can be obtained by analytic continuation 
$t\rightarrow i\chi, \theta \rightarrow {{\pi\over {2}} + i\tau}$ and 
$\beta_i \rightarrow i\beta_i$ of the $R$-charged 
black holes constructed in \cite{bcs,cg,gh}. In literature, there are 
other 
analytic continuations of black holes in adS space giving rise to magnetic 
flux branes, see for example \cite{lpv}}: \begin{equation}
ds^2 =  H^{-2/3}f d\chi^2 + H^{1/3}(f^{-1} dr^2 - r^2 d\tau^2 + r^2
\cosh^2\tau d\Omega_2^2).
\label{bubmet}
\end{equation}
where
\begin{equation}
H_i = 1 - q_i/r^2,
\end{equation}
\begin{equation}
H = {\prod_{i=1}^3}(1-{q_i\over r^2}),
\end{equation}
\begin{equation}
f=1-{\mu\over r^2} + {r^2\over l^2} H.
\end{equation}
The three gauge field potentials $A_{\mu}^i$ are of the form
\begin{equation}
{A_\chi^i} = {{\tilde{q}_i} \over {r^2 - q_i}} - {{\tilde{q}_i} \over 
{r_+^2 - q_i}},
\label{bubgau}
\end{equation}
\noindent where $q_i$ and $\tilde{q}_i$ are given by
\begin{eqnarray}
q_i = \mu \sin^2\beta_i, \qquad\tilde{q}_i=\mu \sin\beta_i \cos\beta_i
\label{charge}
\end{eqnarray}
and $r_+$ is given by the largest positive root of the equation
\begin{equation}
1 - {\mu\over{r^2}} + {r^2\over l^2}\Big(1 -{q_1\over r^2}\Big)
\Big(1-{q_2\over r^2}\Big)\Big(1-{q_3\over r^2}\Big)=0.
\label{hor}
\end{equation}
Note that we have added a constant term in the gauge potential such that 
it vanishes at $r = r_+$. Furthermore, we notice from (\ref{bubgau}) that 
the gauge fields become singular at $r = \sqrt q_i$. However, if we choose 
\begin{equation}
\mu \ge q_i, 
\label{con}
\end{equation}
the singularity occurs at values of $r$ which is always less than 
$r_+$. This is indeed the case as can be seen from (\ref{charge}).
We also have three scalers $X_i$ associated with the configuration. 
They are given by:
\begin{equation}
X_i = H_i^{-1} H^{1\over 3}- (H_i^{-1}H^{1\over 3})_{\rm at~r=r_+}.
\end{equation}
We have again arranged the scalars in such a way that the vanish at the 
$r = r_+$.
We first note that the spacetime is nonsingular in the region $r_+ \le r 
\le \infty$ if $\chi$ has the periodicity
\begin{equation}
\Delta\chi = {{2\pi l^2 r_+^2\sqrt{{\prod_{i=1}^3}(r_+^2 - q_i)}}\over
{2r_+^6 + r_+^4(l^2 - \sum_{i=1}^3 q_i)+\prod_{i=1}^3q_i}}.
\label{per}
\end{equation}
\noindent
As $H$  goes to unity at large $r$, by scaling $l^2/r^2 $, the boundary 
metric can be written as
\begin{equation}
{ds_\Sigma}^2 =d{\chi}^2 + l^2 (- d{\tau}^2 + \cosh^2\tau d\Omega_2^2)
\end{equation}
This is $S^1 
\times dS^3$. The radius of the circle is given by ${\Delta\chi\over 
{2\pi}}$.
On the other hand, at $r=r_+$, the circle, parametrised by $\chi$, 
collapses. However, the two sphere approaches a finite size $r_+^2 
\cosh^2\tau$. This solution, therefore, corresponds to a bubble of radius 
$r_+$ of $d=5,~  {\cal N} = 8$ gauged supergravity. Note that the above 
configuration reduces to the adS-Schwarzschild bubble of \cite{br}
for $q_i =0$\footnote{We note here that even though the R-charged adS 
black holes can have supersymmetric limit (where mass and the charges are 
related in a specific way), the analytically continued bubbles are 
inherently non-supersymmetric. Supersymmetry is broken here due to the 
antiperiodic bounday condition on the fermions along $\chi$.} .

\noindent
 Now we will study  the bubble solution for three distinct cases {\bf:} 
(1) $q_1 = q,~ q_2 = q_3 = 0$, (2) $q_1 ,~  q_2 \ne 0 ,~ q_3 = 0$ and (3) 
$q_1,~q_2,~ q_3 \ne 0 $. 

\noindent {\bf Case 1: }$q_1 = q,~ q_1 = q_2 = 0$

\noindent
Defining $k_1 = 1 - q/l^2$, we get the  radius of the single charged bubble 
is given by
\begin{equation}
r_+^2 = -{k_1l^2\over 2} + {l\over 2}\sqrt{k_1^2 l^2 + 4 \mu}
\end{equation}
Since $ \mu \ge q $,  $ r_+  \ge \sqrt{q}$ 
and the periodicity is
\begin{equation}
{\Delta\chi}_1 = {{2\pi l^2 \sqrt{(r_+^2 -q)}} \over
{2r_+^2 + (l^2 -q)}}.
\end{equation}
For later use, we now discuss the nature of $\Delta\chi_1$ as a function 
of the bubble radius
$r_+$. First consider $ q \le l^2 $.  $\Delta\chi_1$ starts from zero at 
$r_+ = \sqrt q$ and goes to zero for large $r_+$. It has a maximum at 
\begin{equation}
r_+ = r_c = {\sqrt{{l^2 + 3q}\over 2}},
\end{equation}
with 
\begin{equation}
\Delta\chi_{1c} = {\pi l^2\over{\sqrt{2 (l^2 + q)}}}.
\end{equation}
The behaviour of $\Delta\chi_1$ is shown in the following figure. 
\begin{figure}[ht]
\epsfxsize=8cm
\centerline{\epsfbox{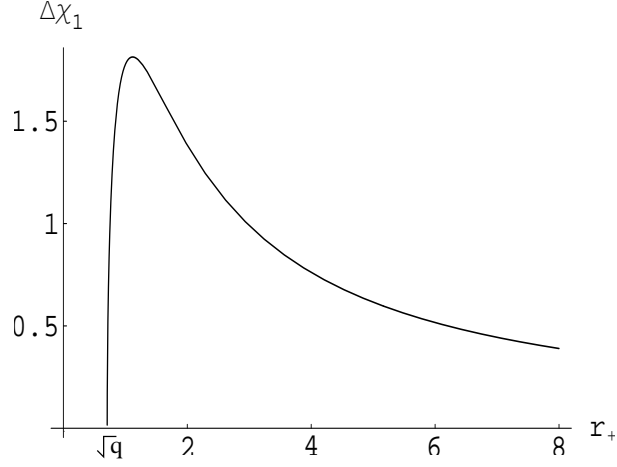}}
\caption{Plot of $\Delta\chi_1$ vs. for $r_+$ $l= 1$, and $ q= 0.5 $}
\end{figure} \\
\noindent
If $q > l^2$, denominator of $\Delta\chi_1$ changes sign at
$ r_+ < \sqrt{q}$,  but this is not allowed value. So the behaviour of $\Delta\chi_1$ is same for all cases.

\noindent {\bf Case 2:} $q_1,~  q_2 \ne  0 ,~  q_3 =0$

\noindent
As before, we define  $ k_2 = 1-(q_1 
  + q_2)/l^2 $, and get the radius of the double charged bubble. This is 
\begin{equation} 
 r_+^2 = - \frac{k_2 l^2}{2} + {\frac{l}{2}}\sqrt{k_2^2 l^2 + 4 \mu
   -\frac{4 q_1 q_2}{l^2}}.
\end{equation}
Since $ \mu \ge q_i $  then $ r_+^2$ is always greater or equal to
largest ${q_i}$. 
The periodicity  is given by
\begin{equation}
{\Delta\chi}_2 ={ {2\pi l^2\sqrt{ (r_+^2 -q_1)(r_+^2 - q_2)}} \over
{2r_+^3 +r_+ (l^2 -q_1 -q_2)}}.
\end{equation}

\noindent
As for $\Delta\chi_2$, when  $ (q_1 + q_2) \le l^2$,   we get the two 
different branches
of periodicity where for small $r_+$, $\Delta\chi_2$ starts from infinity
and goes zero at $r_+= \sqrt{q_2}$ (for $ q_2 < q_1$) and the 
second branch starts
from $r_+=\sqrt{q_1}$ goes to  maximum value and then reaches to zero
for large $r_+$. Here we give the plot for the case where $q_1=q_2=q$.
\begin{figure}[ht]
\epsfxsize=8cm
\centerline{\epsfbox{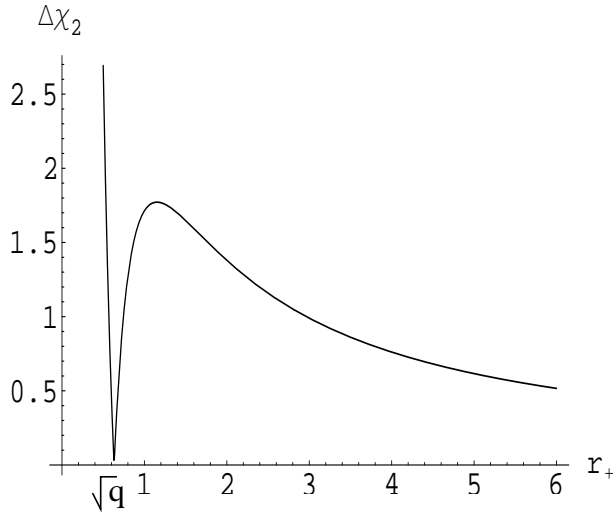}}
\caption{Plot of $\Delta\chi_2$ vs. $r_+$, for $l=1$, and $ q=.4$}
\end{figure} \\
The zero in the plot is at $r_+=\sqrt{q}$ and there is a maximum at
\begin{equation} 
r_+ =r_c= \sqrt{{l^2\over 4} + q + {l\over 4} \sqrt{l^2+16q}},
\end{equation}
with
\begin{equation}
\Delta\chi_{2c} = {\pi l^2{(l+ \sqrt{l^2 +16
        q})}\over{{[{l^2\over 4} + q+ {l\over 4 }{\sqrt{l^2 + 16
       q }}]^{1\over2}(3l +  \sqrt{l^2 + 16 q})}}}.
\end{equation} 
Note that the region below $r_+ = \sqrt q$ is not an allowed region for 
the bubble space-time. This 
can be seen from (16) as $\mu \ge q$.
On the other hand if $ q_1 + q_2 > l^2$, from (17), we see that 
$\Delta\chi_2$ becomes singular at some value of  $r_+ < \sqrt q_1$. 
This is again not an allowed region. The nature of $\Delta\chi_2$ in the 
allowed region is same as the earlier one.

\noindent{\bf Case 3:} $q_1,~ q_2,~ q_3 \ne 0$

\noindent
Next we consider the bubbles for three non-zero charges. We find
that $\Delta\chi_3$ has three zeros and there are 
two branches of periodicity. The zeroes
and the maxima merge into a single zero and single maximum respectively by 
choosing the equality among all three charges. The nature of $\Delta\chi_3$ 
would then be same as $\Delta\chi_1$.

\noindent
Before we proceed further, we would like to make few comments on the 
stability of R-charged bubble solution. In \cite{witten, afhs}, it was 
argued that the bubble solutions are quantum mechanically unstable while 
being classically stable. As for classical stability, we can explicitly 
show in our case that a massless scalar on this charged bubble 
background does not have a normalisable mode with negative mass square on the 
deSitter factor\footnote{The equation of motion of a massless scalar field 
in this background (with massless excitations along $\chi$) is given by
\begin{equation}
(f r^3){\partial_r}^2 \phi + \partial_r(r^3 f)\partial_r \phi = -M^2 r 
\phi,
\end{equation}
where $f = 1 - \mu r^{-2} + r^2 l^{-2}H$, and $M$ satisfies
$\nabla^2_{dS_3} \phi = M^2 \phi$. Now by studying this  equation at $r 
\rightarrow \infty$ and $r\rightarrow r_+$, one easily checks that there 
is no normalisable mode for $M^2 <0$.}. The quantum stability of these 
bubbles are more difficult to analyse. However, as in \cite{br}, we note 
that these bubbles can be embedded in global adS space. Because of the 
periodicity along $\chi$, one needs to identify surfaces in the 
global adS space. This in turn makes the space non-smooth. This may 
prevent nucleation of additional bubbles.

\noindent
We now compute the energy associated with this charged bubbles.
Using counter term subtraction procedure of \cite{bal,liu}, the boundary 
stress tensor can easily be evaluated.
For generic values of charges, the components of the stress tensor are 
given by: 
\begin{equation}
{T_\chi^\chi}= -\frac{ 3}{16 \pi G_5 l^3}{[ \mu + 
  \frac{ l^2}{4}-\frac{2}{3}( q_1 +q_2+q_3)]},
\label{pres}
\end{equation}
\begin{equation} 
{T_\tau^\tau}=\frac{ 1}{16 \pi G_5 l^3}{[ \mu + 
  \frac{ l^2}{4}-\frac{2}{3}( q_1 +q_2+q_3)]},
\label{ener}
\end{equation}
\begin{equation}
{T_\phi^\phi}={T_\psi^\psi}= \frac{ 1}{16 \pi G_5 l^3}{[ \mu + 
  \frac{ l^2}{4}-\frac{2}{3}( q_1 +q_2+q_3)]}. 
\label{momen}
\end{equation}
These components reduce to adS-Schwarzschild bubble 
if we set the charges $q_i$ to zero. In general,  it also follows that due to the presence of the charges, 
the energy of this 
bubble is larger than the adS-Schwarzschild one. This can be seen explicitly from (\ref{ener}).

\noindent
Let us now focus our attention on another class of metric. These are adS 
orbifolds---a version of BTZ black hole in five dimensions. This class of 
space-time was extensively studied in \cite{banadoso, banadost}. The 
asymptotic boundary of these 
black holes are same as the bubbles that we have been discussing so far. 
The only difference as far as the boundary structures are concerned, 
unlike the bubbles, for 
the black holes, the period of $\Delta \chi$ can take any value. Boundary 
stress tensor of these holes were computed in \cite{cai}. They are given 
by the expressions (\ref{pres})--(\ref{momen}) with $\mu,~ q_i =0$. Let us 
now consider the boundary metric $S^1 \times dS^3$ for large $S^1$
along with some radiation matter given by $A^i$ and $X_i$. 
Since,  for bubbles, $\Delta \chi$ has a maximum critical value, the only 
solutions  that are available in the bulk for large $S^1$ 
(or in other words, for large $\Delta\chi$) are the black 
holes. However, as we shrink  $S^1$ down to lower radius, the R-charged 
bubbles also become available below a critical radius given by the 
maximum of the expression in (\ref{per}).
Infact, the energy associated with the bubble is lower than the black hole 
if 
\begin{equation}
\mu >  {2\over 3}(q_1 + q_2 +q_3).
\label{mq}
\end{equation}
This follows from equation (\ref{ener}).
One would therefore expect a transition in the bulk from the black hole 
to the bubble below a critical radius. 

\noindent
Two questions then immediately arise: (1) What triggers this vacuum to 
vacuum transition(if at all) in string 
theory formulated around the background with boundary metric $S^1 \times 
dS^3$? (2) Through dS/CFT 
correspondence, can we understand this transition in gauge theory? 
Neither of these issues are clear to us \footnote{Some understanding along 
this direction will appear in a forthcoming paper 
\cite{brai}}. However, it is tempting to speculate that in string theory, 
it may arise due to instabilities coming from the winding modes along 
$S^1$. On the other hand, in gauge theory on $dS^3\times S^1$, it may arise 
due to the condensation of an order parameter related to bubble radius in 
the bulk.

\noindent
{\bf{Acknowledgments:}} We wish to thank Balram Rai for explaining us some 
of his unpublished works. We have also benefited from discussions with 
Sumit Das and Ashoke Sen on issues related to this work.

\end{document}